\documentclass[a4paper,11pt]{article}
\usepackage{jheppub}
\pdfoutput=1
\usepackage{enumitem}
\usepackage[activate=true]{microtype}
\usepackage{soul,color}
\usepackage{ifthen}
\usepackage{xspace}
\usepackage{slashed}

\usepackage[T1]{fontenc}
\usepackage[utf8]{inputenc}

\newboolean{@draft}
\setboolean{@draft}{false}
\ifthenelse{\boolean{@draft}}{ \providecommand{\hldraft}[1]{\hl{#1}} }{ \providecommand{\hldraft}[1]{#1} }

\providecommand{\gaga}{\gamma\,\gamma}

\newcommand{\sqrtsnn}{\sqrt{s_{_{\textsc{nn}}}}}

\newcommand{\bbbar}    {\rm {b\bar{b}}}
\newcommand{\ccbar}    {\rm {c\bar{c}}}
\newcommand{\qqbar}    {\rm {q\bar{q}}}

\newcommand{\cm}{c.m.\@\xspace}
\newcommand{\elm}{e.m.\@\xspace}
\newcommand{\LumiInt}{\mathcal{L}_{\rm \tiny{int}}}

\newcommand{\fbinv}{\mbox{\ensuremath{\,\text{fb}^{-1}}}\@\xspace}
\newcommand{\nbinv}{\mbox{\ensuremath{\,\text{nb}^{-1}}}\@\xspace}
\newcommand{\abinv}{\mbox{\ensuremath{\,\text{ab}^{-1}}}\@\xspace}
\newcommand{\ie}{i.e.\@\xspace}
\newcommand{\eg}{e.g.\@\xspace}
\newcommand{\etc}{etc.\xspace}

\ifthenelse{\boolean{@draft}}{
\usepackage{draftwatermark}
\SetWatermarkText{DRAFT}
\SetWatermarkScale{3.5}
}{}

\title{Brazilian Participation in the Next-Generation Collider Experiments}

\author[a]{W. L. Ald\'a J\'unior}
\author[b]{C. A. Bernardes}
\author[a]{D. De Jesus Dami\~ao}
\author[c]{M. Donadelli}
\author[d]{D. E. Martins}
\author[e,a]{G. Gil da Silveira}
\author[f]{C. Hensel}
\author[a]{H. Malbouisson}
\author[f]{A. Massafferri}
\author[a]{E. M. da Costa}
\author[a]{C. Mora Herrera}
\author[d]{I. Nasteva}
\author[d]{M. Rangel}
\author[a]{P. Rebello Teles}
\author[b]{T. R. F. P. Tomei}
\author[a]{A. Vilela Pereira}

\affiliation[a]{
Departamento de Física Nuclear e Altas Energias,\\ Universidade do Estado do Rio de Janeiro (UERJ),\\
Rua São Francisco Xavier, 524, CEP 20550-900, Rio de Janeiro, Brazil}

\affiliation[b]{Universidade Estadual Paulista (Unesp), Núcleo de Computação Científica\\
Rua Dr. Bento Teobaldo Ferraz, 271, 01140-070, Sao Paulo, Brazil}

\affiliation[c]{Instituto de Física, Universidade de São Paulo (USP),\\
Rua do Matão, 1371, CEP 05508-090, São Paulo, Brazil }

\affiliation[d]{Universidade Federal do Rio de Janeiro (UFRJ), Instituto de Física,\\
Caixa Postal 68528, 21941-972 Rio de Janeiro, Brazil}

\affiliation[e]{Instituto de Física, Universidade Federal do Rio Grande do Sul ,\\Av. Bento Gonçalves, 9550, CEP 91501-970, Caixa Postal 15051, Porto Alegre, Brazil}

\affiliation[f]{Centro Brasileiro de Pesquisas Físicas (CBPF),\\
Rua Dr. Xavier Sigaud, 150, CEP 22290-180 Rio de Janeiro, RJ, Brazil}

\emailAdd{walter.luiz.alda@cern.ch}
\emailAdd{cesar.augusto.bernardes@cern.ch}
\emailAdd{dilson.de.jesus.damiao@cern.ch}
\emailAdd{marisilvia.donadelli@cern.ch}
\emailAdd{dan.ernani@gmail.com}
\emailAdd{gustavo.silveira@cern.ch}
\emailAdd{carstenh@cbpf.br}
\emailAdd{helena.malbouisson@cern.ch}
\emailAdd{andre.massafferri.rodrigues@cern.ch}
\emailAdd{eliza@cern.ch}
\emailAdd{clemencia.mora.herrera@cern.ch}
\emailAdd{irina@if.ufrj.br}
\emailAdd{rangel@if.ufrj.br}
\emailAdd{patricia.rebello.teles@cern.ch}
\emailAdd{thiago.tomei@cern.ch}
\emailAdd{antonio.vilela.pereira@cern.ch}


\abstract{
This proposal concerns the participation of the Brazilian High-Energy Physics community in the next-generation collider experiments. 
As the LHC at CERN is getting ready for its third data-taking period, Run 3, with increased centre-of-mass energy at the nominal 14 TeV and upgraded detectors, the global High-Energy Physics community is converging towards the materialisation of new powerful colliders which could explore the characteristics of the Higgs boson and advance the search of physics beyond the Standard  Model. 
This means that in the next couple of years the technical proposals for the experiments that will run in such colliders will be submitted for review, collaborations will be formed and large scale funding will be requested at different stages across the globe.
Our aim is to organise a common effort from the Brazilian HEP community of optimised resources and time, in order to produce a higher impact contribution to our chosen experiment and more broadly in the field of High-Energy Physics.\\[1.0ex]
\textit{Paper submitted to the Latin American Strategy Forum for Research Infrastructure}
\vspace{0.5ex}
}

\begin{document} 
\maketitle
\flushbottom

\section{Scientific Context}
\label{sec:scicont}

High-Energy Physics (HEP) explores the elementary particles, which are the fundamental constituents of matter, and their interactions. Elementary particles are the underlying structure at the inner kernel of matter and, at the same time, play an essential role in the evolution of the Universe. The last century has shown that collider accelerators have been amongst the most powerful tools used to explore the deep structure of matter that  enabled the development of a universal quantum field theory --- the standard model (SM). HEP experiments led to important discoveries that go from the identification of heavy quarks, passing by the discovery of the W$^\pm$ and Z\textsuperscript{0} bosons, up to the breakthrough represented by the recent discovery of the Higgs boson at CERN~\cite{Aad:2012tfa,Chatrchyan:2012xdj}, evidence for the Brout-Englert-Higgs mechanism through which fundamental particles acquire their mass.

The \emph{Large Hadron Collider} (LHC) at CERN~\cite{evans:2008zzb}, the most powerful particle collider in operation, has opened new opportunities to explore the energy frontier of physics, colliding protons (pp) at a centre-of-mass energy of 13~TeV, which allows testing the validity of our understanding of microscopic physics up to distances as small as 10\textsuperscript{$-$18} meters. The LHC, operating on the Swiss-French border near Geneva, Switzerland, hosts four main experiments:
ALICE (A Large Ion Collider Experiment)~\cite{aamodt:2008zz},
ATLAS (A Toroidal LHC ApparatuS)~\cite{Aad:2008zzm},
CMS (Compact Muon Solenoid)~\cite{Chatrchyan:2008aa},
and LHCb (Large Hadron Collider beauty)~\cite{alves:2008zz}. ALICE and LHCb are dedicated experiments to heavy ion collisions and $B$-hadron physics, respectively; ATLAS and CMS are general-purpose experiments that aim for a comprehensive study of physics at the TeV scale. All four experiments are undertaken by large collaboration groups, with thousands of participants, and are held as great examples of successful international endeavours. As of this writing, the LHC has just finished its second operation run (Run 2)~\cite{LHCEndRun2}, having delivered more than 150\fbinv of 13~TeV pp collision data to the experiments. Each of the collaborations has published hundreds of scientific papers, contributing to the steady advance of physics knowledge. 

\begin{figure}
\centering
\includegraphics[width=\textwidth]{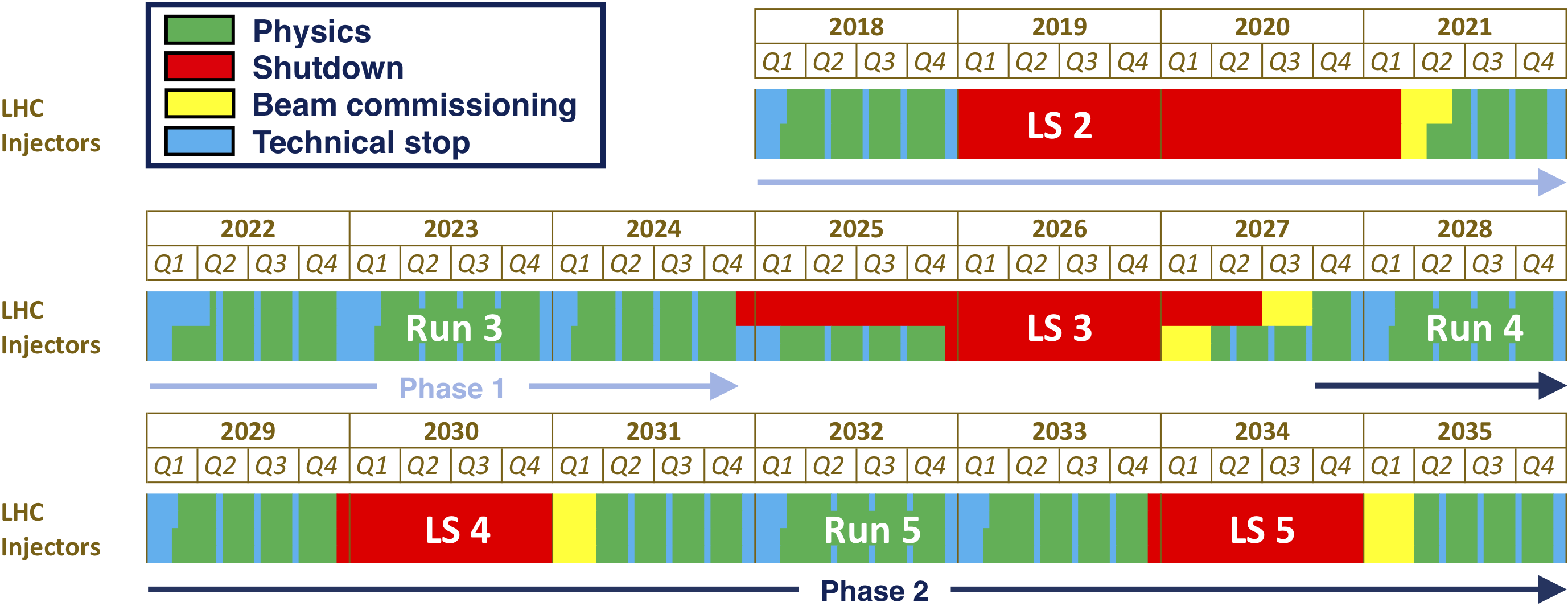}
\caption{Calendar for the operation of the LHC in Phases 1 and 2 up to 2035.}
\label{lhctimeline}
\end{figure}

The design, construction, commissioning and operation of a particle accelerator and experiments is a multi-decade endeavour. The LHC project was originally proposed at the 1984 ECFA-CERN workshop \cite{Asner:1984jv}, approved by the 100th CERN Council Session in 1994, delivered its Conceptual Design Report in 1995 and accelerated its first beams in 2008. The High-Luminosity LHC (HL-LHC) is an upgrade to the accelerator chain that will increase the instantaneous luminosity up to an ultimate performance of 75~Hz/nb. In a similar vein, 
the experiments will undergo their own upgrades to prepare for the HL-LHC era. According to the official LHC schedule, the goal is to start the high luminosity phase in 2027 and deliver 3000\fbinv of pp collision data to ATLAS and CMS until 2035, as detailed in Fig.~\ref{lhctimeline}. It is safe to say that the path of collider-based HEP is charted until that date.

The question on the medium-term future of collider-based HEP is still open. Studies on the design of the next accelerator have already been started and a series of Conceptual Design Reports have been put forward:

\begin{itemize}

\item The Circular Electron Positron Collider (CEPC) \cite{Lou2019} is a project led by IHEP-Beijing, proposing the construction of a 100~km circumference machine near Qinhuangdao. The CEPC centre-of-mass energy is proposed to be 240~GeV, allowing it to work as a Higgs boson factory;

\item The International Linear Collider (ILC) \cite{Behnke:2013xla} and the Compact Linear Collider (CLIC) \cite{Charles:2018vfv} are two electron-positron collider projects that currently coexist under the Linear Collider collaboration umbrella. The ILC centre-of-mass energy is projected to go up to 1~TeV with a 31~km long accelerator, whilst CLIC is projected to reach up to 3~TeV with a 50~km long accelerator. As of this writing, the ILC proposal for construction in the Kitakami mountains, in Japan, is not favoured by the Japanese government, as informed by the Science Council of Japan. The CLIC proposal for construction in Geneva is still to be evaluated;

\item The Future Circular Collider (FCC) \cite{Abada:2019zxq,Benedikt:2018csr,Abada:2019lih,Abada:2019ono} is again an umbrella project that supports three different proposals: a high-energy upgrade for the LHC (HE-LHC), hosted in the same tunnel; and a 100 km circumference tunnel built in the Geneva region, hosting either an electron-positron (FCC-ee) or a hadron collider (FCC-hh).

\end{itemize}

All of those projects envision first beams after 2030.

\section{The Brazilian Community of Accelerator High-Energy Physics}

Brazilian research institutions currently participate in all four LHC experiments. The country also takes part in the PHENIX and STAR experiments at the Relativistic Heavy Ion Collider (RHIC), and had strong participation in the DELPHI experiment at the Large Electron-Position Collider (LEP) and D0 experiment at the Tevatron. Since 2008, the Brazilian HEP community has a coordination board in RENAFAE, the Brazilian Network for High-Energy Physics \cite{renafae}. In its foundation, this network defined the main goals as (i) to promote scientific and technological \hldraft{development} for the investigation of elementary particles and fundamental interactions, (ii) to consolidate the research in High-Energy Physics by extending the technical competences of Brazilian researchers, (iii) to build partnerships with the Brazilian enterprises to research and develop software and instrumentation for international scientific collaboration, and (iv) to coordinate the activities of the Brazilian research groups, especially those in international collaborations. 

{\tolerance=500 
In 2014, an all-encompassing project on Astroparticle and High-Energy Physics, \texttt{AstroHEP.br} was submitted to the Brazilian Research Council (CNPq), with a proposal to create a National Institute of Science and Technology (INCT). The goal of that project was to ``...\textit{facilitate the sharing of resources, seeking synergies between the different groups, greatly increasing the competitiveness of Brazilian groups.}'' Although \texttt{AstroHEP.br} did not come to fruition, its spirit of pooling the effort of the Brazilian community with the goal of bringing it an ever-higher standard of excellence is laudable. Following that spirit, the overall goal of this proposal is to organise the Brazilian community for an effective participation in the next-generation collider experiments.
\par}

Along the years, RENAFAE has been the financial \hldraft{representative} for the High-Energy Collaborations by its link with the Brazilian Scientific agencies, allowing the payment of M\&O and \hldraft{funding} for the LHC experiments. Although efforts to increase financial support from the Brazilian Scientific agencies, RENAFAE has struggled to support broader research projects in the international collaborations and \hldraft{the presence} of Brazilian researchers at CERN, \hldraft{to support} qualified personnel developing long-term activities, especially in instrumentation.

Therefore, the Brazilian HEP community must follow another path to provide results of higher impact in the international scientific community and qualify the body of active researchers in the Brazilian institutions, especially graduate students and post-doctoral researchers. Concrete actions are proposed in this document to achieve these goals in the next decade.

\section{Objectives}

The main objectives of our proposal are:
\begin{enumerate}[label=\alph*)]
\item To organise the community of experimental HEP in Brazil to become a member of only one large experimental collaboration 
in the next high-energy collider, joining forces to contribute to the development, construction, commissioning and operations 
of the detector as a coordinated effort. In particular we would like to avoid very specific short-aim and small-visibility projects 
and favour longer-term contributions that can make a greater impact within the chosen collaboration and in our research field.

\item Organise and connect smaller \hldraft{groups} in the country with an interest in more specific experiments, for example future DIS colliders.

\item Restructure the financial and practical organisation of large HEP collaboration membership, to evolve towards a 
National Institute type of entity (INCT) which could provide independent and guaranteed funding for the development of the projects, 
including the mobility needs for doctoral and technical students as well as researchers, 
and also provide a local affiliation for post-doctoral researchers, engineers and technical professionals.

\item In the long term, we would like to build local physical infrastructure with one or more national laboratory sites, 
both for instrumentation research and development and for computing, data processing and analyses (analogous to current Grid Tier-1). 
 
\end{enumerate}





\section{Current status and expected challenges}

It is fair to state that the Brazilian institutes have been contributing to CERN experiments and any other large-scale collaboration without following a clear strategy. It is common in Europe and in the USA that the financial agencies approve grants for the participation of institutes in large experiments in a long-term manner. This method has been proven to be very efficient providing foundations for a sustained commitment of the institutes. This is far from how Brazil and Latin America financial agencies operate. We believe that is the main challenge faced by the Brazilian institutes to participate on large-scale experiments. 
\hldraft{We believe that this challenge must be the highest priority of this strategy forum.}

As a consequence, there is a huge challenge to perform detector R\&D in the Brazilian institutes due to the lack of infrastructure and engineering capabilities. 
\hldraft{In addition, we are not able to properly train our students in the local institutes.}

Brazilian institutes are not able to attract researchers from around the world. This issue is related to the lack of post-doctoral positions and the current funding strategy. The long-term grant should allow the institutes to hire post-doctoral researchers with \hldraft{competitive} salaries and with the possibility to have them based at CERN. 

Another crucial challenge is the financial support for mission trips to the experiment. This must be treated as an important point since person-power is one of the most important contributions Brazilian institutes give to large-scale experiments. It is important to note that the training of students is mostly done during mission trips. In fact, dedicated mobility projects were created to support these trips to CERN, namely HELEN and EPLANET. 
\hldraft{Unfortunately, since 2015 there has been no similar initiative, which has had a clear impact on the participation of Brazilian students and researchers at CERN.}

RENAFAE was created in 2008 to mitigate the issues described above. It is fair to state that there was some progress in the pursuit of a Brazilian strategy but the current status is that the Brazilian organisational model for particle physics is far from being optimal.

Although the European model is proven to be successful one can not simply copy it to the Brazilian reality. The Brazilian model needs to be developed together with the scientific societies and the government. We propose that RENAFAE must become an institute that will discuss funding directly with the financial agencies. Currently, RENAFAE is associated to CBPF with a low budget and simple structure. It is desirable that RENAFAE becomes an institute with dedicated researchers and personnel in order to sustain the top priority scientific activities for the next future collider.

Even if RENAFAE does not become an institute, we urge for long-term grants. A national scientific strategy with clear identification of the high-priority large-scale scientific goals for Brazilian institutes is necessary. A profound modification of the Brazilian financial agencies rules must be proposed in order to maintain and improve the participation of Brazilian institutes at the forefront of particle physics. As simple examples, we propose the creation of post-doctoral fellowships to be based in the experiment and dedicated funding for mission trips for students as an extension of the current {\textit{sandwich programme}}, similar to the HELEN/EPLANET programmes.



\newpage
\section{Operational costs}

Assuming the Brazilian community in a single experiment is $\sim$50 authors, and the experimental collaborations comprise $\sim$3000 authors, the Brazilian ``fair-share'' contribution is 1.7\%.

Considering a landscape to consolidate the research and development of software and instrumentation, the goal for financial support for the next decade would aim a total of 100~researchers with regular contributions for M\&O and detector upgrades. 
\hldraft{Moreover, the R\&D of instrumentation and detector technologies must be in part carried out in a common infrastructure for better collaboration among Brazilian researchers and higher impact.}
Thus, the creation of a dedicated laboratory similar to an INCT-like initiative is a prime goal for the next decade.

\section{Computing requirements}

\begin{figure}
\centering
\includegraphics[width=\textwidth]{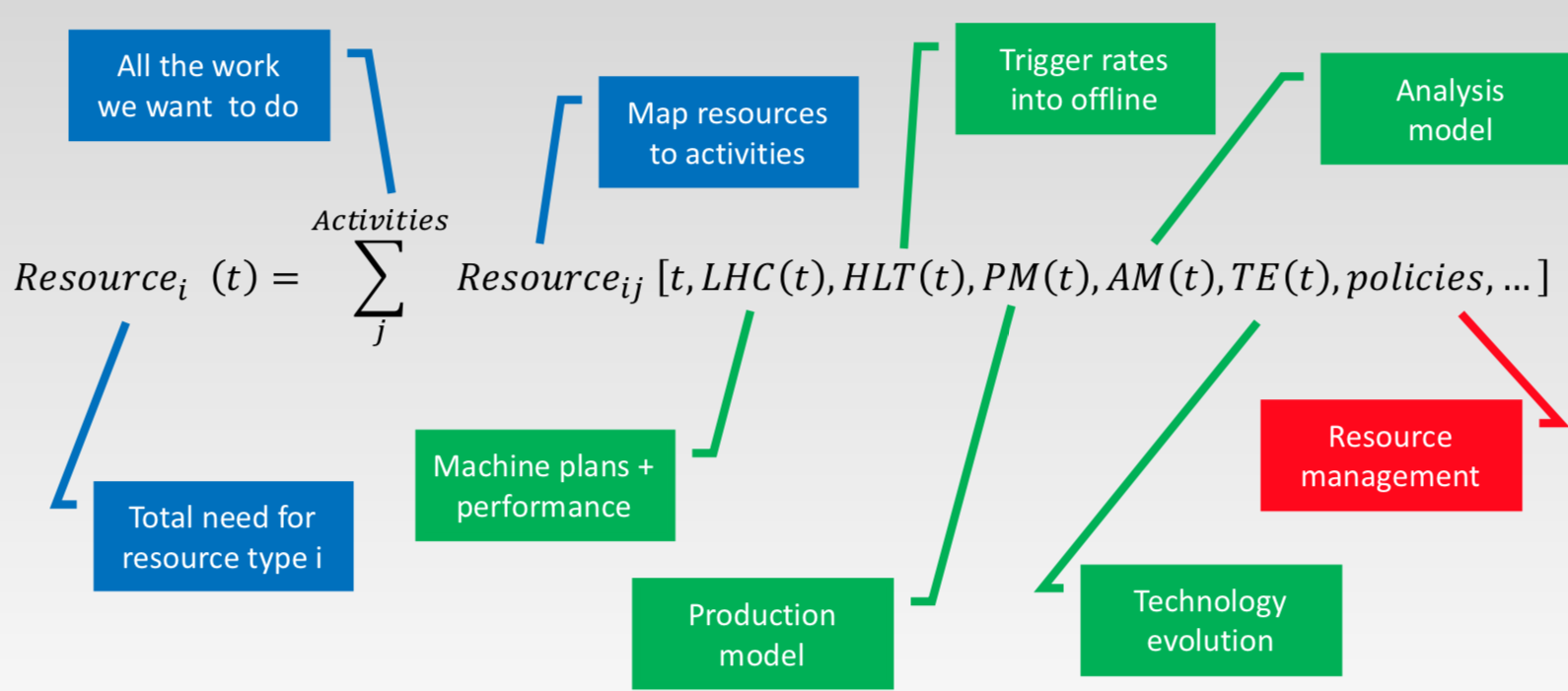}
\caption{CMS computing model. Extracted from \cite{DLangeCHEP2019}.}
\label{fig:CMS-computing-model}
\end{figure}

The \emph{computing model} of a HEP experiment estimates the computing requirements of a given resource (CPU power, disk, tape, \etc) as a function of:
\begin{itemize}
    \item the experiment's physics programme,
    \item the accelerator plans and performance,
    \item the rate and amount of data acquired by the experiment,
    \item the production model -- the rate and amount of simulated collision events to be produced by the collaboration,
    \item the analysis model -- the way collaborators access and run over the data,
    \item the expected technology evolution,
    \item and the policies for usage and management of that given resource.
\end{itemize}
Figure~\ref{fig:CMS-computing-model} exemplifies the CMS computing model as of the LHC Run 2; that model is flexible enough to be adapted to the future accelerators' case.
As of this writing, the only inputs that can be estimated with any accuracy are the accelerator plans and the experiment's data acquisition design.
The experiment's production model, analysis model and resource policies can, at best, be supposed to be similar to those of the LHC.
Therefore, in what follows, we will simply quote the data acquisition design rates as a proxy for the computing requirements of the experiment.
For comparison, the CMS HL-LHC proposal envisions a $\sim 60 \text{ GB/s}$ data rate~\cite{Collaboration:2283193}.
On the other hand, the expected technology evolution until the start of the new accelerator experiments merits some discussion.

\subsection{Expected technology evolution}

``\textit{During the last years HEP used the assumption of a $\sim$20\% improvement rate for CPU and disk resources\ldots{} It looks like the $\sim$20\% number is too optimistic and needs to be revised.}''~\cite{berndp-s}.
Even assuming a lower improvement rate of 10\% year-by-year, that translates to at least a factor of 2 improvement in the price-performance ratio of computing resources by 2030.
We also draw attention to the fact that \emph{heterogeneous computing} is quickly becoming a reality at the LHC experiments, with CPU-GPU hybrid systems being proposed for deployment already during Run~3~\cite{CMS-DP-2018-059,patatrack}.
Other alternatives like CPU-FPGA and CPU-GPU-FPGA are also being studied.
It stands to reason that usage of hybrid systems will be commonplace at the next accelerators.

\subsection{Computing at CLIC and ILC}

For both the CLIC and ILC scenarios, the computing requirements are comparable to those of the regular LHC experiments, \ie{} much smaller than those of the high-luminosity upgrade.
ILC operation foresees 1 millisecond bunch trains at 5 Hz, whilst CLIC foresees 156 nanosecond bunch trains at 50 Hz.
That translates to a data rate to offline storage of $\sim$1 GB/s for the ILC and $\sim$10 GB/s for CLIC.
We therefore assume that the overall computing requirements of the experiment will be smaller than those of the HL-LHC.

\subsection{Computing at CEPC}

The computing requirements at CEPC, on the other hand, will be similar to those at the HL-LHC. The CEPC CDR quotes only that ``\textit{With the level-1 trigger operating at 100 kHz, the total raw data rate is 2 TBytes/s.}"
This corresponds to 16~Tb/s for the L1 data rate, roughly 40\% of the values proposed for the CMS HL-LHC proposal.
We assume a similar ratio for the data rate to offline storage, leading to a figure of $\sim$24 GB/s.
We therefore assume that the overall computing requirements of the experiment will be comparable to those of the HL-LHC.

\subsection{Computing at FCC}

The computing requirements for the FCC-ee are comparable to those of the CEPC, being very similar machines.
On the other hand, the requirements for an FCC-hh detector are formidable, and only preliminary calculations are presented in the CDR.
The figures quoted are, for the L1 data rate, 250 TBytes/s for calorimetry and muon systems (about 10 times more than in the ATLAS and CMS Phase-II scenario), and 1--2 PB/s from the tracker.
We assume a similar ratio for the data rate to offline storage, leading to a figure of $\sim$600 GB/s.
We therefore assume that the computing requirements of the FCC-ee experiment will be comparable to those of the HL-LHC, whilst for the FCC-hh experiment they will be significantly larger.

\section{Physics motivation}

\subsection{Heavy ions}

The study of QCD in high density and temperature conditions
using heavy ion collisions will receive
a vital contribution from upgrades in the LHC and
future colliders. As described in
Refs.~\cite{Citron:2018lsq,Dainese:2016gch}, the gain comes in several sectors of heavy ion physics
as a consequence of the increase in the centre-of-mass
energy per nucleon pair and integrated luminosity. The centre-of-mass energy used in the heavy ion collisions
will depend on the accelerator properties designed for proton-proton collisions.
The relation between the centre-of-mass energy in proton-proton
collisions ($\sqrt{s}$) and centre-of-mass energy per
nucleon-nucleon collision ($\sqrt{s_{\mathrm{NN}}}$) is
given by: $\sqrt{s_{\mathrm{NN}}}=\sqrt{s}\sqrt{Z_1Z_2/A_1A_2}$,
where $Z$ is the number of protons and
$A$ is the number of protons plus neutrons inside the nucleus~\cite{Citron:2018lsq}.

In the HE-LHC, a centre-of-mass energy of $\sqrt{s}=27~\mathrm{TeV}$ for
proton-proton collisions corresponds to $\sqrt{s_{\mathrm{NN}}}=10.6~\mathrm{TeV}$ for
PbPb collisions ($Z=82$, $A=208$) and $17~\mathrm{TeV}$ for pPb collisions.
The integrated luminosity for a data-taking period of
one month is expected to be a factor of two larger than current LHC runs, \ie{} of the order of 6~\nbinv per experiment.
The possibility to use smaller nuclei compared to Pb (\eg{} $^{40}Ar$ or $^{129}Xe$)
can increase the $\sqrt{s_{\mathrm{NN}}}$ and the integrated luminosity. For
example, the integrated luminosity for $XeXe$ collisions at $\sqrt{s_{\mathrm{NN}}}=11.5~\mathrm{TeV}$
could be around a factor 2--3 larger than in PbPb collisions~\cite{Citron:2018lsq}.

In the FCC case, a centre-of-mass energy of $\sqrt{s}=100~\mathrm{TeV}$ for
proton-proton collisions corresponds to $\sqrt{s_{\mathrm{NN}}}=39~\mathrm{TeV}$ for
PbPb collisions and $63~\mathrm{TeV}$ for pPb collisions. The
integrated luminosity for a one-month run in FCC is estimated to be around
one order of magnitude higher than in current LHC runs, \ie{} ${\approx}$ 33~\nbinv.
\hldraft{Similar estimates are valid for the hadronic collider CEPC-SppC}, 
with a slightly lower centre-of-mass energy $\sqrt{s_{\mathrm{NN}}}=20-30~\mathrm{TeV}$~\cite{Dainese:2016gch}.

The physics motivations for HE-LHC and FCC are similar, with a clear
advantage for FCC in terms of studies of hard probes and kinematics
coverage. Three main general physics motivations are described in Refs.~\cite{Citron:2018lsq,Dainese:2016gch}:

\begin{itemize}

\item \textit{Novel access to QCD thermodynamics and QCD equilibration processes}: at HE-LHC and FCC energies, a
heavy ion collision results in a strongly-interacting medium (quark-gluon plasma, QGP)
significantly denser and hotter, expanding for a longer period of time and larger volume compared to current LHC energies.
This feature results in stronger collective signatures. In addition, the QGP will reach
temperatures of the order of $1~\mathrm{GeV}$, where charm quarks will start to participate as a thermal degree of freedom  in the equation of state describing the medium.

\item \textit{Characterisation of dense QCD matter through hard-scattering processes}:
with the significant increase in the centre-of-mass energy, \hldraft{hard probe production will considerably increase}, with the possibility to study boosted top quarks
and also possibly the Higgs boson production~\cite{dEnterria:2015mgr,Apolinario:2017sob,dEnterria:2017jyt}. The studies related
to the production of boosted top quarks will allow to access time evolution information of the QGP medium and the role of color coherence~\cite{Apolinario:2017sob}.
In addition, it is expected to \hldraft{reach sensitivity for the identification} of secondary production of charm quarks
from interactions of quarks and gluons constituents of the QGP.

\item \textit{Exploration of saturated parton densities in a previously-uncharted, ultra-dense kinematic domain}:
at HE-LHC and FCC energies, \hldraft{very low values of ($x, Q^2$), 
where parton saturation is expected to be detected, will be accessed.} 
For example, \hldraft{in pPb collisions $x$ values of ${\sim}5\times10^{-6}$ in the Pb nucleus, for rapidity $y\approx5$, could be reached.}
Very important complementary studies can be done in electron-hadron (ion) colliders, \eg{} LHeC,
to explore the partonic structure of nucleons and nuclei at small-$x$.

\end{itemize}

\subsection{Flavour physics}

The large statistics and/or the clean experimental environment will provide the next generation of collider experiments a great potential to study flavour physics. 
\hldraft{The understanding} of \hldraft{electroweak} penguin mediated decays will provide important model-discrimination \hldraft{for the remaining flavour anomalies}.
Standard Model stringent tests can be done by studying charged lepton flavour violation, lepton flavour
universality violation and heavy-flavoured neutral meson oscillation.

Currently, the measurements agree with the lepton flavour universality to a precision of $0.15\%$ \cite{Lusiani:2018ced}, with the uncertainty dominated by the measurement of the $\tau$ leptonic properties. This uncertainty can be reduced in future colliders with a larger data sample.
A similar conclusion can be reached with respect to the current \hldraft{flavour} anomalies~\cite{Aaij:2014ora,Aaij:2015oid,Abdesselam:2016llu}.

It is clear that flavour physics will remain an important pillar of high-energy physics in the future collider experiments.
\hldraft{It is an area that has long received important contributions from Brazilian researchers and institutes.}

\subsection{Exclusive two-photon production of the Higgs boson in ultraperipheral proton and nuclear collisions}

Heavy ions accelerated at high energies are surrounded by huge electromagnetic (\elm) fields generated by the collective action of their $Z$ individual proton charges. In the equivalent photon approximation (EPA), such strong \elm{} fields can be identified as quasireal photon beams with very low virtualities $Q^{2} < 1/R_A^{2}$ and large longitudinal energies of up to $\omega_{\rm max}\approx\gamma_L/R_A$, where $R_A$ is the radius of the charge and $\gamma_L=E_{\rm beam}/m_{N,p}$ is the beam Lorentz factor for nucleon or proton mass $m_{N,p} = 0.9315,\,0.9382$~GeV.

On the one hand, since the photon flux scales as the squared charge of each colliding particle, $Z^2$, photon-photon cross sections are enhanced millions of times for heavy ions (up to $Z^4 \approx 5\cdot 10^{7}$ for Pb-Pb) compared to proton or electron beams. On the other, proton (and lighter ions) feature larger $\omega_{\rm max}$ values thanks to their lower radii $R_A$ and larger beam $\gamma_L$ factors, and can thereby reach higher photon-photon center-of-mass (\cm) energies. At the LHC energies, photons emitted from nuclei (with radii $R_A\approx 1.2\,A^{1/3}$~fm) are almost on-shell (virtuality $Q<$~0.06~GeV, for mass numbers $A>$~16), and reach longitudinal energies of up to hundreds of GeV, whereas photon fluxes from protons ($R_A\approx$~0.7~fm) have larger virtualities, $Q\approx$~0.28~GeV, and longitudinal energies in the TeV range. 

Table~\ref{tab:1} summarizes the relevant characteristics of photon-photon collisions in ultraperipheral collisions (UPCs) of proton and nuclear beams \hldraft{at the HL-LHC, HE-LHC and FCC}~\cite{dEnterria:2019jty}.

\begin{table}
\centering
\caption[]{Summary of the characteristics of photon-photon collisions in ultraperipheral proton and nuclear collisions at the HL-LHC, HE-LHC, and FCC: 
(i) Nucleon-nucleon \cm\ energy $\sqrtsnn$, (ii) integrated luminosity per run $\LumiInt$, 
(iii) beam energies $\rm E_{beam}$, (iv) Lorentz factor $\gamma_L$, (v) effective charge radius $R_A$, 
(vi) photon ``maximum'' energy $\omega_{\rm max}$ in the \cm\ frame, and
(vii) ``maximum'' photon-photon \cm\ energy $\sqrt{s_{\gaga}^{\rm max}}$.
The last two columns list the $\rm \gaga\to H$ cross sections and the 
expected number of Higgs events for the quoted $\LumiInt$ per system.}
\label{tab:1}
\vspace{0.2cm}
\resizebox{\textwidth}{!}{%
\begin{tabular}{lccccccccc} \hline
System\!\!  & \!\!$\sqrtsnn$\!\! & \!\!$\LumiInt$\!\! & \!\!\!\!$E_{\rm beam1}+E_{\rm beam2}$\!\! & $\gamma_L$ &  $R_A$ & $\omega_{\rm max}$ 
       & $\sqrt{s_{\gaga}^{\rm max}}$ & $\rm \sigma(\gamma\gamma\to H)$ & $\rm N(\gamma\gamma\to H)$ \\
\hline
Pb-Pb & 5.5  TeV &  10~nb$^{-1}$ & 2.75 + 2.75 TeV & 2950 & 7.1 fm &  80 GeV & 160 GeV & 15~pb & 0.15 \\ 
Xe-Xe & 5.86 TeV &  30~nb$^{-1}$ & 2.93 + 2.93 TeV & 3150 & 6.1 fm & 100 GeV & 200 GeV &  7~pb & 0.21 \\ 
Kr-Kr & 6.46 TeV & 120~nb$^{-1}$ & 3.23 + 3.23 TeV & 3470 & 5.1 fm & 136 GeV & 272 GeV &  3~pb & 0.36 \\ 
Ar-Ar & 6.3  TeV & 1.1~pb$^{-1}$ & 3.15 + 3.15 TeV & 3400 & 4.1 fm & 165 GeV & 330 GeV & 0.36~pb & 0.40 \\ 
O-O   & 7.0  TeV & 3.0~pb$^{-1}$ & 3.5 + 3.5 TeV   & 3750 & 3.1 fm & 240 GeV & 490 GeV & 35~fb & 0.11 \\ 
p-Pb  & 8.8 TeV  &   1~pb$^{-1}$ & 7.0 + 2.75 TeV  & 7450, 2950 & 0.7, 7.1 fm & 2.45 TeV, 130 GeV & 2.6 TeV & 0.17~pb & 0.17 \\ 
p-p   & 14 TeV   &   1~fb$^{-1}$ & 7.0 + 7.0 TeV   & 7450 & 0.7 fm & 2.45 TeV& 4.5 TeV & 0.18~fb & 0.18 \\ \hline 

Pb-Pb & 10.6 TeV &  10~nb$^{-1}$ & 5.3 + 5.3 TeV   & 5700 & 7.1 fm & 160 GeV & 320 GeV & 150~pb & 1.5 \\ 
Xe-Xe & 11.5 TeV &  30~nb$^{-1}$ & 5.75 + 5.75 TeV & 6200 & 6.1 fm & 200 GeV & 400 GeV &  60~pb & 1.8 \\ 
Kr-Kr & 12.5 TeV & 120~nb$^{-1}$ & 6.25 + 6.25 TeV & 6700 & 5.1 fm & 260 GeV & 530 GeV &  20~pb & 2.4 \\ 
Ar-Ar & 12.1 TeV & 1.1~pb$^{-1}$ & 6.05 + 6.05 TeV & 6500 & 4.1 fm & 320 GeV & 640 GeV & 1.7~pb & 1.9 \\ 
O-O   & 13.5 TeV & 3.0~pb$^{-1}$ & 6.75 + 6.75 TeV & 7300 & 3.1 fm & 470 GeV & 940 GeV & 0.11~pb & 0.33 \\ 
p-Pb  & 18.8 TeV &   1~pb$^{-1}$ & 13.5 + 5.3 TeV  & 14\,400, 5700 & 0.7, 7.1 fm & 4.1 TeV, 160 GeV & 4.2 TeV & 0.45~pb & 0.45 \\ 
p-p   & 27 TeV   &   1~fb$^{-1}$ & 13.5 + 13.5 TeV & 14\,400 & 0.7 fm & 4.1 TeV & 8.2 TeV & 0.30~fb & 0.30 \\ \hline 

Pb-Pb &  39 TeV  & 110~nb$^{-1}$ & 19.5 + 19.5 TeV & 21\,000 & 7.1 fm & 600 GeV & 1.2 TeV & 1.8~nb & 200 \\ 
p-Pb  &  63 TeV  &  29~pb$^{-1}$ & 50. + 19.5 TeV  & 53\,300, 21\,000 & 0.7,7.1 fm & 15.2 TeV, 600 GeV & 15.8 TeV &  1.5~pb & 45 \\ 
p-p   & 100 TeV  &   1~fb$^{-1}$ & 50. + 50. TeV   & 53\,300 & 0.7 fm & 15.2 TeV &  30.5 TeV & 0.70~fb & 0.70 \\\hline
\end{tabular}
}
\end{table}

The observation of the Higgs boson in UPCs, via its dominant $\bbbar$ decay channel, relies on the measurement of 
two exclusive b-jets with invariant masses peaked at $m_{\rm H}$, on top of a background of 
$\gamma\gamma \rightarrow \bbbar, \ccbar, \qqbar$ continuum pairs, where charm and light (q = u,\,d,\,s) quarks 
are misidentified as b-quarks. By assuming realistic jet acceptance, reconstruction performances, and (mis)tagging efficiencies, and applying appropriate kinematical selections on the jet $p_T$ and angles, it has been shown that the H$(\bbbar)$ signal can be reconstructed on top of the $\gaga\to\bbbar, \ccbar, \qqbar$ continuum backgrounds~\cite{dEnterria:2019jty}. 

Reaching 3$\sigma$ evidence of UPC Higgs-production at HL-LHC and at HE-LHC requires factors of about $\times$200 and $\times$30 more integrated luminosities in Ar-Ar and Kr-Kr collisions, respectively, than currently planned for both machines. Factors of ten in integrated luminosity can be gained running for the duration ($10^7$ s) typical of a proton-proton run, rather than the nominal 1-month heavy-ion operation. This would open up the possibility of a $3\sigma$ evidence at the HE-LHC, but would still fall too short for any feasible measurement at the HL-LHC. The measurement of ${\rm \gaga \rightarrow H} \rightarrow \bbbar$ would yield about 20 (5) signal counts after all selections in Pb-Pb (p-Pb) collisions for their nominal integrated luminosities per run. 

Observation of the photon-fusion Higgs production at the $5\sigma$-level is achievable in the first FCC run by 
combining the measurements of two experiments (or doubling the luminosity in a single one) in Pb-Pb, 
and by running for about 8 months (or 4 months and combining two experiments) in the p-Pb case.
The feasibility studies presented here indicate the Higgs physics potential open to study in $\gamma\gamma$ ultraperipheral 
ion collisions at current and future CERN hadron colliders, eventually providing an independent measurement of the 
H-$\gamma$ coupling not based on Higgs decays but on a $s$-channel production mode, as well as of its total 
width combining the photon-fusion measurement and the H\,$\to\gaga$  decay branching ratio~\cite{dEnterria:2019jty}.

\subsection{Higgs boson pair production}

The Higgs boson discovery by the ATLAS and CMS experiments was conducted with only a few percent of the entire dataset foreseen for the LHC. The first two data taking periods of the LHC allowed the measurement of the Higgs boson production cross sections and decay branching fractions in various channels with increasing precision~\cite{Aad:2015zhl,Aad:2015mxa,Khachatryan:2014kca,Aaboud:2018zhk,Aaboud:2018urx,Sirunyan:2018kst}. Despite all these measurements, the Higgs scalar potential is still largely unconstrained, including the Higgs boson trilinear self-coupling $\lambda_{HHH}$.  Measurements of the strengths of the Higgs boson self-couplings and their comparison to Standard Model predictions are necessary to verify the electroweak symmetry breaking mechanism of the Standard Model. The measurement of the Higgs potential is therefore a high priority goal on the physics programme of all future colliders~\cite{Cepeda:2019klc,deBlas:2019rxi}. The existence of an extended scalar sector or the presence of new dynamics at higher scales could modify the Higgs boson self-couplings. 

Higgs boson pair production ($HH$) is a probe of the  Higgs boson trilinear self-coupling $\lambda_{HHH}$ and it can be assessed at hadron colliders, where the production is dominated by gluon fusion, $gg \rightarrow  HH$, and at lepton colliders via double Higgs-strahlung, $e^{+}e^{-} \rightarrow ZHH$, (relevant at low energies), or via vector boson fusion (VBF),  $e^{+}e^{-} \rightarrow  HH \nu_{e} \bar{\nu}_{e}$, (for centre-of-mass energies of 1 TeV and above)~\cite{Cepeda:2019klc,deBlas:2019rxi}. A particular advantage of $e^{+}e^{-}$ colliders is that the Higgs self-coupling can also be assessed using single Higgs production processes. 

At the LHC, $HH$ non-resonant and resonant searches have been conducted by ATLAS and CMS in various decay modes with datasets collected at 13 TeV, with the $HH \rightarrow b\bar{b}b\bar{b}$, the $HH \rightarrow b\bar{b} \gamma \gamma$ and $HH \rightarrow b\bar{b} \tau^{+} \tau^{-}$ driving the sensitivity to the SM $HH$ production~\cite{Aad:2019uzh,Sirunyan:2018two}. Besides the direct determination of the Higgs self-coupling through the study of Higgs boson pair production, an indirect measurement is also possible exploring the NLO electroweak corrections to single Higgs measurements. In this direction, ATLAS has recently determined the first experimental constraint on $\kappa_{\lambda} = \lambda_{HHH}/\lambda^{SM}_{HHH}$ from single Higgs boson analyses~\cite{ATL-PHYS-PUB-2019-009}. 

The four next generation $e^{+}e^{-}$ colliders under consideration intend to provide precision measurements of Higgs interactions at high energy. However, only colliders that reach a centre-of-mass energy of 500 GeV have access to $HH$ production. 
In the following, the prospects for studies at the future colliders are outlined.

\begin{itemize}
    \item \textit{HL-LHC}: despite its small production cross-section,  more than 10$^{5}$ $HH$ pairs per experiment are expected to be produced at the HL-LHC. Prospects of the future measurements expect that a  combined significance of 4$\sigma$ can be achieved with all systematic uncertainties included, considering the production of $HH$ ($HH \rightarrow b\bar{b}b\bar{b}$, the $HH \rightarrow b\bar{b} \gamma \gamma$,  $HH \rightarrow b\bar{b} \tau^{+} \tau^{-}$, $HH \rightarrow b\bar{b} VV (l l \nu \nu)$, $HH \rightarrow b\bar{b} ZZ$ ) through gluon fusion. However, these studies consider the current Run-2 analyses techniques and measurements extrapolated to the centre-of-mass energy of 14 TeV and integrated luminosity of 3\abinv, to estimate the response of the upgraded ATLAS and CMS detectors at the HL-LHC assuming an average pileup of 200 collisions per bunch crossing~\cite{Cepeda:2019klc}. 
    
    \hldraft{It is important} to note that the VBF production mode will profit  from the extended tracker acceptance and, consequently, the improved ability to identify forward jets from the hard-scattering interaction, which will yield to an increased background rejection benefiting the various decay modes targeted by VBF analyses. 
    
    \item \textit{CEPC, FCC-ee, CLIC, ILC}: the ECFA Higgs@Future Colliders working group has performed fits that use the expected  measurement accuracies for the various single Higgs observables to assess the sensitivity to deviations of the Higgs self-couplings from its SM values~\cite{deBlas:2019rxi}. Results are given for a 1-parameter fit to the SM plus a varying $\kappa_{\lambda}$ (for CEPC, FCC-ee, CLIC, ILC) and for a fit that includes the possibility of other new physics effects modeled by the SMEFT (for FCC-ee and ILC). 
    
    Studies of $HH$  production cross-section as a function of centre-of-mass energy  were performed for $ZZHH$ and $HH \nu_{e} \bar{\nu}_{e}$ processess with unpolarised beams~\cite{Roloff:2019crr}. Despite the smallness of the production cross-sections,  working with polarised beams, as planned for linear $e^{+}e^{-}$ colliders, can raise the cross-section of the $HH \nu_{e} \bar{\nu}_{e}$ reaction significantly, by almost a factor of 2~\cite{Roloff:2019crr}. 
     The $HH \nu_{e} \bar{\nu}_{e}$ reaction requires energies above 1 TeV, as reported by analyses done with full simulation using detailed detector models by the CLICdp~\cite{Roloff:2019crr} and ILC~\cite{Tian:2013qmi} groups, respectively.  
     
    Interesting to note is the opposite dependence of the cross-sections for $HH$ production as a function of $\kappa_{\lambda}$  at lepton colliders. Double Higgs-strahlung, $e^{+}e^{-} \rightarrow ZHH$ reaction gives stronger constraints on positive deviations ($\kappa_{\lambda} >$ 1), while VBF   $HH \nu_{e} \bar{\nu}_{e}$ is better in constraining negative deviations ($\kappa_{\lambda} <$ 1)~\cite{deBlas:2019rxi}. 
    
    At the ILC,  $e^{+}e^{-} \rightarrow ZZHH$ is foreseen with 4\abinv of data collected at 500 GeV~\cite{Tian:2013qmi}. 
    The CLIC program foresees the study of the Higgs self-coupling at 1.5 and 3~TeV stages, with integrated luminosities of 2.5\abinv and 5\abinv, respectively. With the studies performed by the CLICdp group at 1.4~TeV, evidence for $\nu \bar{\nu} HH$ production is found with 3.6$\sigma$ and the $ZHH$ process observed with a significance of 5.9 $\sigma$.
\end{itemize}

\subsection{New Physics at TeV scale}

\hldraft{Searches for evidences of New Physics beyond the SM have been performed at the LHC at the TeV scale.} Part of these searches are focused in simplified models with effective Lagrangians~\cite{Alves:2011wf}, where several constraints can be tested in the LHC energy regime. Considering the upcoming high-luminosity regime of the LHC, such models are well suited to probe New Physics sensitivity and test different production processes, like initial- and final-state radiation. 
Other possibilities \hldraft{include} different approaches for the strong dynamics~\cite{Appelquist:2016viq} and the coupling of the Higgs boson to new particles in the strong and electroweak sectors~\cite{Jana:2017hqg}.

Searches for candidates for Dark Matter are one of the main topics under study with the LHC data, such as WIMPs and neutralinos. In order to properly describe the relic abundance at the freeze-out, most effective models aim for large mass mediators, \hldraft{for which} thermal averaged cross \hldraft{sections} may be reachable at the LHC energies. The main strategy for searching for signals of Dark Matter is the determination of the transverse missing energy ($\slashed{E}_{T}$)~\cite{Fox:2011pm} in events with initial-state radiation. Such events are characterised by the presence of a SM signal associated to large $\slashed{E}_{T}$, \eg{} mono-jets~\cite{Chatrchyan:2011nd,Chatrchyan:2012me,Khachatryan:2014rra,Aad:2011xw}, mono-photon~\cite{Khachatryan:2014rwa,Sirunyan:2017ewk}, mono-Higgs~\cite{CMS:2017ysu,Sirunyan:2017hnk,Sirunyan:2018fpy,Sirunyan:2018gdw}, mono-$Z$~\cite{Khachatryan:2015bbl,Khachatryan:2016mdm,Sirunyan:2017hci,Sirunyan:2017onm}, and monotop~\cite{Khachatryan:2014uma,Sirunyan:2019gfm} or in association with quarks tops \cite{Sirunyan:2018dub,Sirunyan:2018gka}. 
\hldraft{The current constraints are for the DM mediator mass 1.6--1.8~TeV and for the DM mass 0.4--0.7~TeV.} 
The high-luminosity regime will open the possibility to enhance such studies given the large amount of data, investigate all mono-X channels and provide insight about the electroweak sectors of Dark Matter \cite{Aboubrahim:2019vjl}. 
\hldraft{Furthermore, $e^+e^-$ collisions at CLIC will reach a centre-of-mass energy of 3~TeV and collect an amount of data of 2\abinv, a scenario which allows extended DM searches}~\cite{deBlas:2018mhx}.

%
%

\bibliography{refs}
\bibliographystyle{JHEP}
\end{document}